\documentclass[prl,twocolumn,showpacs,preprintnumbers,superscriptaddress,amsmath,amssymb]{revtex4}

\usepackage{epsf,color,graphicx}

\usepackage[dvips]{epsfig}

\begin{document}

%\preprint{}

\title{Microcavity polariton light emitting diode}



\author{Daniele Bajoni}
\author{Elizaveta Semenova }
\author{Aristide Lema\^{i}tre}
\author{Sophie Bouchoule}
\author{Esther Wertz}
\author{Pascale Senellart}
\author{Jacqueline Bloch}\email[]{jacqueline.bloch@lpn.cnrs.fr}
\affiliation{CNRS-Laboratoire de Photonique et Nanostructures, Route
de Nozay, 91460 Marcoussis, France}

\date{\today}

\begin{abstract}
Cavity polaritons have been shown these last years to exhibit a rich
variety of non-linear behaviors which could be used in new polariton
based devices. Operation in the strong coupling regime under
electrical injection remains a key step toward a practical polariton
device. We report here on the realization of a polariton based light
emitting diode using a GaAs microcavity with doped Bragg mirrors.
Both photocurrent and electroluminescence spectra are governed by
cavity polaritons up to 100 K.
\end{abstract}

\pacs{71.36.+c, 78.60.Fi, 73.50.Pz, 78.55.Cr}
%\keywords{}

\maketitle

 As first predicted by Purcell in 1946 \cite{Purcell46}, spontaneous
emission of light can be strongly modified when inserting an emitter
in a resonant cavity. When the emitter is located at an antinode of
the electromagnetic field, the emission of light can be strongly
accelerated or even become reversible if the light-matter coupling
is strong enough. In such strong coupling regime, the degeneracy
between the emitter and the photon mode is lifted giving rise to two
light-matter entangled eigenstates spectrally separated by the Rabi
splitting. This strong coupling regime has been first evidenced for
atoms in ultra-high finesse microcavities \cite{Thompson92} and more
recently in various solid state systems like a superconducting
q-bit\cite{Wallraff2004}, quantum well excitons in inorganic
\cite{Weisbuch92} or organic semiconductors \cite{Lidzey98},
intersubband transition in two dimensional electron gas \cite{Dini}
or excitons in single quantum dots
\cite{Yoshie,Reithmaier,Peter2005}. Cavity polaritons, resulting
from the strong coupling regime between excitons in quantum wells
and cavity photons, have been the subject of intensive research
since their discovery in 1992 \cite{Weisbuch92}. In two dimensional
cavities, each exciton with a given in-plane wave vector $k_{//}$ is
coupled to the photon mode with the same $k_{//}$. The strong
coupling regime gives rise to two polariton branches with a
pronounced energy trap of the lower branch close to $k_{//}$ = 0
\cite{Houdre94}. The bosonic nature of polaritons in addition to the
strong polariton-polariton interactions are responsible for a rich
variety of non-linear behaviours in inorganic materials: polariton
accumulation in the ground state and formation of a macroscopically
occupied coherent state
\cite{Deng2002,Kasprzak2006,Christopoulos2007,Snoke2007,BajoniPRL2007},
photon pair emission \cite{Romanelli2007} or spin current generation
\cite{Leyder2007}. These features are promising for future quantum
applications but have up to now only been observed under optical
pumping. Electrical injection of these entangled light-matter states
is a key step toward the implementation of practical, compact
devices. So far electrical injection of cavity polaritons has been
reported only in organic semiconductors \cite{Tischler} where the
coupling strength is very large (Rabi splitting of several hundreds
of meV). More recently indications of electrical injection of
polaritons using intersubband transitions in GaAs have been claimed
\cite{Sapienza2007}. However to our knowledge non-linearities have
not been observed using either of these schemes. The GaAs system,
when relying on interband transitions, is very attractive since its
strong non-linearities could be exploited in an electrically pumped
device.

In the present letter, we report on the realization of a GaAs
microcavity containing quantum wells and surrounded by two doped
Bragg mirrors. Both photocurrent spectra and electroluminescence
evidence two well-resolved polariton branches up to 100 K.
Increasing the current density, electroluminescence spectra reveal
the progressive transition from the strong coupling regime to the
weak coupling regime because of exciton screening at high carrier
density. These findings open the way toward future polariton based
devices.

\begin{figure*}[]
\includegraphics[width= 0.8\textwidth]{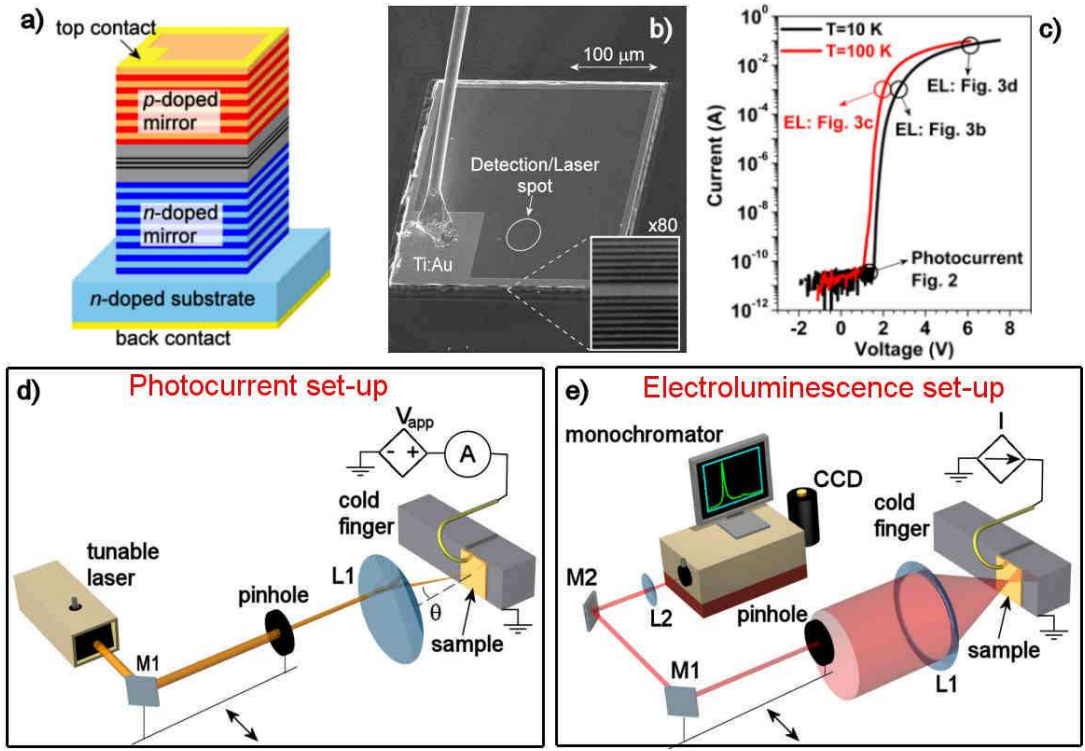}
\caption{(Color online)a) Schematic view of the sample structure. b)
Scanning electronic microscope image of a mesa showing the lateral
contact and the top injection wire. Also indicated is the typical
size of the optical excitation/detection spot. In the inset a
magnified image of the mesa edge reveals the multilayer structure
with the cavity in the middle. c) Current-voltage characteristics
measured at 10 K and at 100 K. The circles indicate the operating
conditions used in the experiments described in this paper. d)
Schematic view of the set-up for photocurrent measurements: the
sample is excited with a tunable laser controlling both the energy
and the angle of incidence of the exciting beam while the induced
current is detected. e) Schematic view of the set-up for
electroluminescence measurements: a current is injected through the
sample and the induced emission is monitored for various detection
angles.}\label{Fig1}
\end{figure*}

The cavity structure is shown on fig. 1a: it consists in an undoped
GaAs cavity containing three $In_{0.05}Ga_{0.95}As$ quantum wells
and surrounded by two $Ga_{0.9}Al_{0.1}As/Ga_{0.1}Al_{0.9}As$ Bragg
mirrors. The mirrors are doped to inject or collect carriers into or
from the active region. The cavity layer is undoped so that the
excitonic character of the quantum well emission is preserved. The
sample was grown by molecular beam epitaxy on an n-doped GaAs
substrate.  The top (resp. bottom) mirror is p-doped (resp. n-doped)
and contains 20 (resp. 24) pairs. The p-doped Bragg mirror is
completed by a highly-doped p++ GaAs thin top contact layer. Al
graded concentrations at each interface in the Bragg mirror are
introduced to optimize mirror resistance. The structure is designed
so that the quantum well exciton emission is close to resonance with
the cavity mode at low temperature. A slight wedge of the cavity
layer allows fine tuning of the cavity mode.

Square mesas of 300 $\mu m$ lateral size were etched down to the
GaAs substrate using optical lithography and wet chemical etching.
Ti-Au was evaporated to form the top lateral p-type contact, and was
completed in the inside by a semitransparent (40 percents
transmission) Au layer to ensure uniform electrical injection over
the entire diode surface. AuGeNi was evaporated and alloyed on the
backside of the wafer to form the bottom n-type contact. Such mesas
were defined at different positions of the wafer, corresponding to
different detunings between the cavity mode and the quantum well
energy.

The sample is kept at low temperature in a cold finger cryostat. For
photocurrent measurements (fig.1d), the sample is biased with a
stabilized voltage source. A cw Ti:Saph laser beam is focused on the
sample (50 $\mu m$ diameter spot) with a 50 mm focal lens (L1) . The
excitation incident angle is controlled by a pinhole placed close to
the Fourier plane of L1 (1$^{\circ}$ angular resolution). For
electroluminescence measurements (fig.1e), the inverse scheme is
used. The sample is driven by a stabilized current source. The
emission is collected through L1 and through the pinhole and
analyzed with a monochromator followed by a nitrogen cooled Si CCD
camera.

The electrical characteristics of the device at 10K and 100 K are
shown in fig. 1c. A typical diode-like behavior is evidenced with a
band alignment around 1.5 V. Below 1.5 V, the built-in electric
field prevents forward current from flowing through the structure.
Because of the low operating temperature, reverse current is not
detectable. Under optical excitation, photogenerated carriers flow
from the quantum well toward the contacts (electrons into the n
contact and holes into the p contact) thus generating a reverse
photocurrent. On the other hand, when a voltage above 1.5 V is
applied, a forward current is generated through the structure:
electrons and holes recombine in the undoped region giving rise to
an electroluminescence signal. Both operating conditions are
described below.

Let us first consider the reverse bias regime \cite{Fisher95} where
a voltage of 1.3V is applied to the photodiode at 10K. The
photocurrent spectra shown in fig.2a were measured by optically
exciting the sample at various incident angles Two dips are
evidenced in each photocurrent spectrum. For a given incident angle
the laser selectively excites polariton states with $k_{//} =
E(k_{//}) sin(\theta)/\hbar c$, where E(k$_{//}$) is the polariton
energy. Thus these photocurrent measurements directly map the
polariton dispersion by probing the absorption. The energy of the
photocurrent dips is summarized in fig.2b as a function of $\theta$:
two polariton branches are observed presenting the anticrossing
characteristic of the strong coupling regime. These dispersion
relations are well reproduced considering the radiative coupling
between each exciton of given k$_{//}$ and energy $E_{x}$ and the
photon mode of same k$_{//}$ and energy $E_{c}=\frac{\hbar
c}{n}\sqrt{(\frac{2\pi}{L_{c}})^{2}+k_{//}^{2}}$ , where n is the
effective refractive index of the cavity and $L_{c}$ the cavity
layer thickness. The best fit is obtained using a Rabi splitting of
5 meV and a detuning   $\delta = E_{c}(0)-E_{x}(0) = -1 meV$.

\begin{figure}[]
\includegraphics[width= \columnwidth]{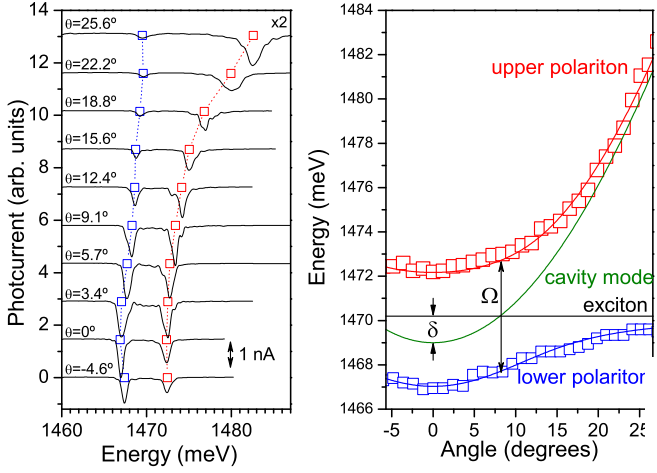}
\caption{(Color online)a) Photocurrent spectra measured for various
detection angles. T = 10K, the applied voltage is V = 1.3 V at the
limit for reverse polarization; b) Squares: energy of the
photocurrent peaks as a function of the excitation angle. Continuous
lines: calculated energy of the uncoupled cavity mode and exciton
line and of the upper and lower polariton branch using a Rabi
splitting   = 5 meV and $\delta$  = -1 meV.}\label{Fig2}
\end{figure}

\begin{figure}[b]
\includegraphics[width= 0.5\textwidth]{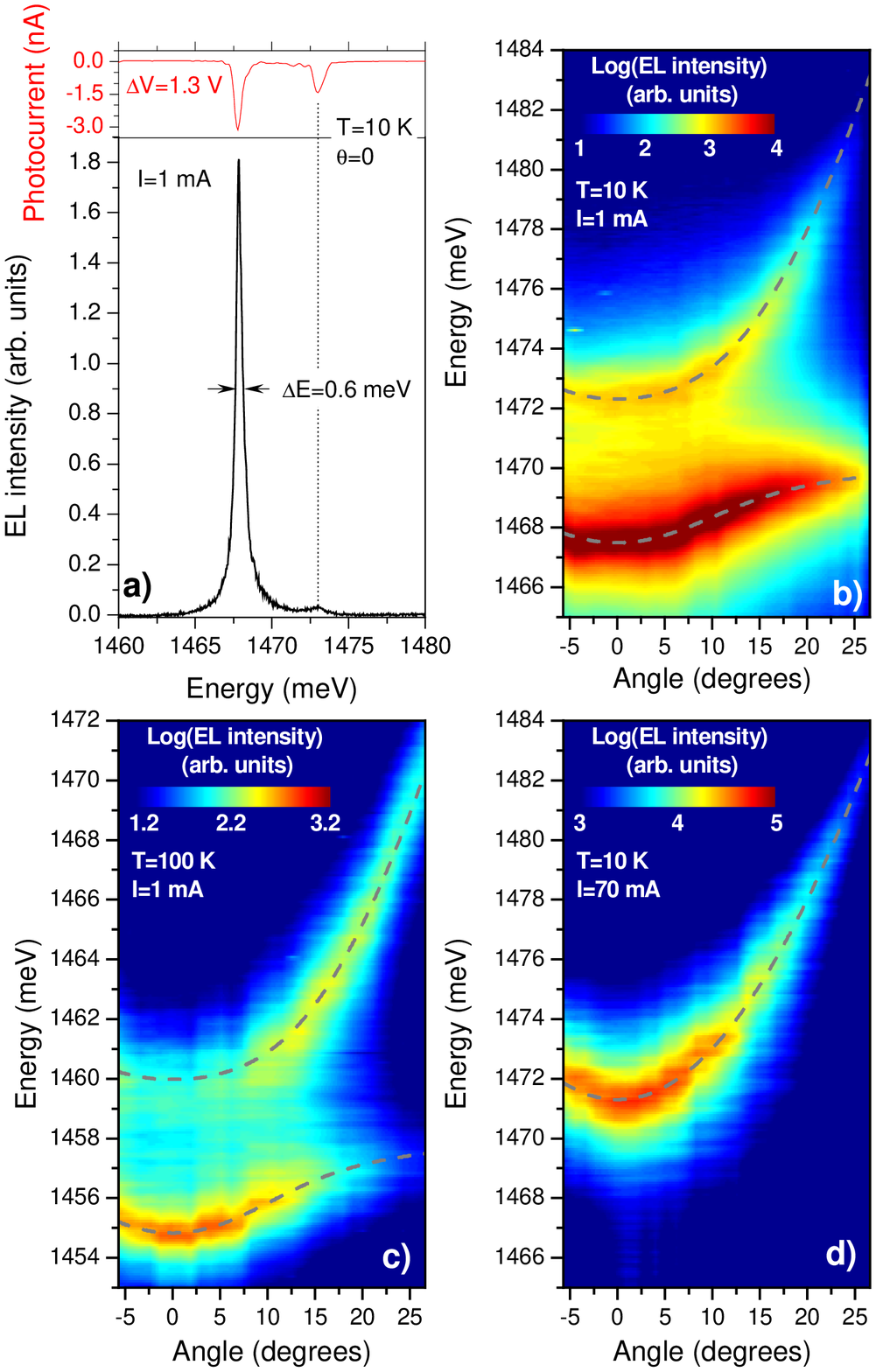}
\caption{(Color online)a) Photocurrent spectrum measured under
normal incidence of the exciting laser beam for V = 1.3 V, and
electroluminescence spectrum measured under normal incidence for I =
1 mA. b)-d) Electroluminescence spectra measured as a function of
the detection angle for b) I = 1 mA and T = 10 K, c) I = 1 mA and T
= 100 K, d) I = 70 mA and T = 10K. In b) and c), the dotted lines
correspond to the calculated upper and lower polariton branches
using $\Omega$  = 5 meV and the intensities are in logscale. In d)
the dotted line corresponds to the cavity mode
dispersion.}\label{Fig3}
\end{figure}

These measurements demonstrate that the exciton photon strong
coupling regime can be achieved in a GaAs based vertical cavity with
doped Bragg mirrors. Nevertheless they do not imply the ability to
electrically inject polaritons. When the diode is driven by a
forward current, the strong coupling regime could for example be
screened by the electric field induced by charge accumulation.

To test the polariton emission properties of our sample, we now
consider the forward bias regime where electron-hole pairs are
electrically injected into the photodiode undoped region. Fig.3a
shows an electroluminescence spectrum measured under normal
incidence for an applied bias V = 2.5 V. Emission peaks are exactly
at the same energy as the photocurrent dips measured at the same
point under normal incidence. This shows that Stark shift is
negligible in the considered bias conditions\cite{Fisher95}. The
emission linewidth is 0.9 meV and 0.6 meV for the upper and lower
polariton branches emphasizing the good optical quality of this
doped microcavity sample. Evidence of the strong coupling regime is
obtained by monitoring the electroluminescence spectra as a function
of the emission angle. Fig.3b summarizes the emission spectra
measured under moderate injection conditions (current intensity I =
1 mA, which corresponds to a current density j $\simeq$0.01
A/cm$^{2}$, assuming uniform electrical injection). Both polariton
branches are clearly evidenced with the characteristic s-shape of
the lower branch. The total exciton-polariton density $n_{x}$ can be
estimated, assuming that all the injected carriers form excitonic
polaritons and then recombine. This hypothesis is supported by the
linear dependence of the total electroluminescence intensity with
current intensity, indicating that non-radiative processes can be
neglected, together with the absence of GaAs emission. $n_{x}$ is
then simply given by : $n_{x} = j \tau /e$, where e is the electron
charge and $\tau$ the polariton recombination time averaged over the
lower branch (typically 200-400 ps in such samples
\cite{Sermage96}). For I = 1 mA, the estimated value of the
polariton density is of the order of $n_{x} \simeq 1-3\cdot 10^{9}
cm^{-2}$. Fig. 3d summarizes the measured electroluminescence
spectra at a higher current intensity (I = 70 mA, j $\simeq
0.78\quad A/cm^{2}$): in this case, a single emission line is
observed in the electroluminescence spectra with the characteristic
dispersion relation of the bare cavity mode. The estimated carrier
density for I = 70 mA lies around 7 $10^{10} cm^{-2}$, and is beyond
the Mott density (typically of the order of some $10^{10} cm^{-2}$)
\cite{Kappei2005}. In this density regime, excitons are bleached by
the screening of the Coulomb interaction and phase space filling:
the optical properties are therefore dominated by the recombination
of uncorrelated electron-hole pairs. This continuum of unbound
electron-hole pairs is in the weak coupling regime with the cavity
mode, and the emission is simply filtered by the cavity mode. Thus
fig. 3b and fig.3d illustrate the transition from the strong
coupling to the weak coupling regime when increasing the density of
electrically injected electrons and holes.

To investigate the operability range of the present polariton
device, electroluminescence measurements were performed at different
temperatures. The detuning between the exciton and the cavity mode
increases with temperature. The present wafer exhibits detuning
close to zero only up to 100 K. This has determined the upper limit
of the operating temperature that could be tested in the present
sample. As shown in fig. 3c, the strong coupling regime still
persists at 100K: both polariton branches are observed with a
significant broadening of the emission lines (0.8 meV and 1.5 meV
for the lower and upper branch) induced by increased
polariton-phonon interaction.

To summarize, we report the first clear demonstration of
electrically driven polariton emission in a semiconductor
microcavity. This opens the way toward new sample designs suitable
for the realization of electrically pumped polariton lasers. Note
that in the present work, the excitation is non resonant so that the
relaxation bottleneck \cite{Tassone97} limits the occupation of the
ground state. To circumvent this problem and reach quantum
degeneracy under electrical injection, cavities with reduced
dimensionality can be considered \cite{Snoke2007,BajoniPRL2007}.
Another approach could be the use of resonant tunnelling to directly
inject polaritons into the lower branch states. Finally strong
technological developments\cite{grandjean} could possibly allow the
same approach in the GaN system \cite{Christopoulos2007,Semond2005}
to reach room temperature polariton lasing.

This work was funded by the european project ``Clermont 2"
(MRTN-CT-2003-503677), by "C'nano Ile de France" and "Conseil
G\'en\'eral de l'Essonne".

\end{document}